# Subharmonic lock-in detection and its optimisation for femtosecond noise correlation spectroscopy

M. A. Weiss[1], F. S. Herbst[1], S. Eggert[1], M. Nakajima[2], A. Leitenstorfer[1], S. T. B. Goennenwein[1] & *T. Kurihara[1,3]

[1]Department of Physics, University of Konstanz, D-78457 Konstanz, Germany.

[2]Institute of Laser Engineering, Osaka University, 565-0871 Osaka, Japan.

[3]The Institute for Solid State Physics, The University of Tokyo, 277-8581 Kashiwa, Japan.

*Corresponding author: takayuki.kurihara@issp.u-tokyo.ac.jp

**Although often viewed as detrimental, fluctuations carry valuable information about the physical system from which they emerge. Femtosecond noise correlation spectroscopy (FemNoC) has recently been established to probe the ultrafast fluctuation dynamics of thermally populated magnons by measurement of their amplitude autocorrelation. Subharmonic lock-in detection is the key technique in this method, allowing to extract the pulse-to-pulse polarisation fluctuations of two femtosecond optical pulse trains transmitted through a magnetic sample. Here, we present a thorough technical description of the subharmonic demodulation technique and of the FemNoC measurement system. We mathematically model the data acquisition process and identify the essential parameters which critically influence the signal-to-noise ratio of the signals. Comparing the model calculations to real datasets allows validating the predicted parameter dependences and provides a means to optimise FemNoC experiments.**

## I. INTRODUCTION

Fluctuations in condensed matter systems contain rich information about their microscopic interactions. In equilibrium, spontaneous fluctuations are connected to the susceptibility[1,2]. Nevertheless, measuring their dynamical properties in the time domain provides access to the fundamental mechanisms of quantum electromagnetic fields[3,4] as well as the thermodynamical nature of solid-state systems[5–10]. In spin systems, spin noise spectroscopy (SNS) is a widely established tool to study the paramagnetic spin fluctuations in the megahertz to gigahertz regime[11]. In SNS, the spin noise is imprinted onto the polarisation state of a continuous-wave laser by the Faraday effect. Using SNS based on ultrashort laser pulses, the investigation of spin fluctuations with long coherence times in paramagnets was reported[8–10,12]. However, fluctuations of the magnon systems in exchange-coupled solids often exhibit picosecond dephasing times and, in particular in antiferromagnets, natural resonance frequencies up to several terahertz – well beyond the reach of conventional SNS.

In the last decade, experimental approaches granting access to fluctuation dynamics in the terahertz frequency range were developed in the emerging field of subcycle quantum optics[3,4,13,14]. The common principle of these techniques relies on probing the polarisation noise encoded e.g., by stochastic fluctuations of the vacuum electric field in a high-repetition rate (tens to hundreds of MHz) femtosecond probe pulse train. In practice, a variety of methodologies exist, which grant access to the encoded information: In one of the early examples of subcycle quantum optics (ref.[3]) the ultrafast stochastic properties of mid-infrared vacuum fluctuations are probed by analysing the polarisation state histogram of a single probe-pulse train detected via electro-optic sampling. While this approach enables femtosecond time resolution, dynamical information of the vacuum field remains unresolved. To access the femtosecond dynamics, a different method, which we denote as femtosecond noise correlation spectroscopy (FemNoC), was proposed in ref.[4]. Here, the polarisation noise imprinted by the vacuum electric field on two femtosecond optical probes separated in time and transmitted through an electro-optic crystal are detected using two individual polarimetric detectors. By real-time multiplication and averaging of the output signals, the autocorrelation function of the vacuum fluctuations is directly revealed at ultrafast time scales. Recently, we combined this FemNoC technology with SNS and applied it to the investigation of magnetisation noise in the orthoferrite $Sm_{0.7}Er_{0.3}FeO_3$, thereby successfully resolving the incoherent noise dynamics of antiferromagnetic magnons with femtosecond resolution[15].

The crucial ingredient for the implementation of FemNoC experiments is the extraction of the pulse-to-pulse polarisation fluctuations that carry the information about the stochastic properties of the investigated system. Since a femtosecond laser system operating at high repetition rates is used in these measurements, the detection of the pulse-to-pulse fluctuations requires a high-speed electronic data acquisition scheme. For this purpose, subharmonic lock-in detection[16] is frequently used. Here, the electronic signal output from the polarimetric detector is demodulated at half the frequency of the laser repetition rate in a radio-frequency (RF) lock-in amplifier to extract only the noise component from the predominating signals at the fundamental repetition rate. Sub-harmonic demodulation-based FemNoC offers an array of advantages, namely access to broad detection bandwidths and a high dynamic range due to suppression of the repetition-rate frequency component. While the FemNoC scheme was shown to successfully retrieve the pulse-to-pulse fluctuations, the parameters relevant for an efficient subharmonic detection have not been systematically studied. Indeed, a detailed technical description and mathematical formulation of the data acquisition procedure has not been put forward.

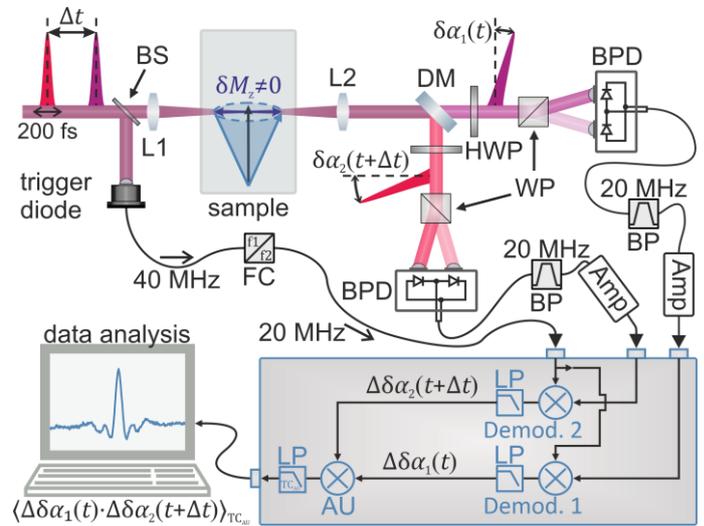

Figure 1: Schematic illustration of the experimental setup and electronic data processing. Two spectrally separated 200 fs pulses (red and purple) of variable time delay $\Delta t$ are focussed onto a magnetically ordered sample and, upon transmission, experience a polarisation rotation proportional to the out-of-plane spin fluctuations $\delta M_z$ due to the Faraday effect. The corresponding rotation angles $\delta\alpha_{1,2}$ are measured in separate polarimeters. The detector outputs are bandpass filtered and amplified before being sent into a radio-frequency lock-in amplifier (grey box). In the lock-in amplifier, the pulse-to-pulse fluctuations from each branch are extracted using subharmonic demodulation. The two demodulation outputs are then multiplied and averaged in the arithmetic unit (AU) of the lock-in amplifier, yielding the cross-correlation function $\langle\delta\alpha_1\delta\alpha_2\rangle$ as a function of the delay time $\Delta t$. BS: beam splitter; FC: frequency converter; L1: microscope objective lens; L2: collimation lens; DM: dichroic mirror; HWP: half-wave plate; WP: Wollaston prism; BPD: balanced photodetector; BP: electronic bandpass filter; Amp: transimpedance amplifier; Demod.: demodulator; LP: digital low-pass filter; AU: arithmetic unit.

Here, we provide a full mathematical formulation of the subharmonic lock-in detection for the FemNoC scheme, providing analytical solutions allowing to predict the optimum combination of parameters for the FemNoC signals. We compare experimental datasets and simulation results to confirm the validity of our model. Parameters that critically influence the signal-to-noise ratio are identified, allowing for significant sensitivity enhancement of the FemNoC signals.

## II. EXPERIMENTAL METHOD AND SETUP

### a. Overview of the optical FemNoC setup

Our experimental setup rests on the following principle: Two linearly polarised, spectrally distinct femtosecond optical probes are transmitted through the magnetic sample (see Figure 1). Here, by the magneto-optical Faraday effect, the transient out-of-plane magnetisation noise imprints as polarisation fluctuations on two femtosecond pulse trains transmitted through the sample. The probes are separated using a dichroic mirror (DM) and the induced polarisation noise is then detected with two separate polarimetric detectors. These detectors yield an output signal periodic with the repetition rate $f_{rep}$ of the laser pulse train. We deploy a radio-frequency (RF) lock-in amplifier (UHFLI, Zurich Instruments[17]) to extract the fluctuation of the detector output by subharmonic demodulation. This step yields the pulse-to-pulse polarisation fluctuations $\Delta\delta\alpha_1(t)$ and $\Delta\delta\alpha_2(t+\Delta t)$ of the two pulse trains and at the same time, it removes any static background $\alpha'$ as well as spurious $\frac{1}{f}$-like noise components arising in the detection electronics. In the next step, $\Delta\delta\alpha_1(t)$ and $\Delta\delta\alpha_2(t+\Delta t)$ are multiplied in real time within the RF lock-in amplifier, and the resulting product is averaged subsequently. This procedure results in the correlated polarisation noise $\langle\Delta\delta\alpha_1(t)\cdot\Delta\delta\alpha_2(t+\Delta t)\rangle$. At the same time, the uncorrelated background noise (shot noise of the optical probe, Johnson noise in the electronics, etc.) arising in the separate polarimetric detection arms is eliminated. By changing the variable time delay $\Delta t$ of the pulse trains, the correlation function $\langle\delta M_z(t)\cdot\delta M_z(t+\Delta t)\rangle$ of the transient magnetisation fluctuations is retrieved on femtosecond time scales.

For our experiment, we exploit a modelocked Er:fibre laser emitting pulses with a temporal width of 150 fs, a central wavelength of 1.55 µm and a total energy of 5 nJ at a repetition rate of 40 MHz. This output is first frequency doubled in a periodically poled lithium niobate (PPLN) crystal before being spectrally and spatially separated by a dichroic mirror. The resulting optical probe pulses are linearly polarised and have central wavelengths of 767 nm and 775 nm with 3 to 4 nm bandwidths, respectively. One of the pulse trains is sent over an optical delay line, where it is temporally shifted by a variable time-delay $\Delta t$ with respect to the other femtosecond pulse train. We use a tension-free transmissive microscope objective lens with a numerical aperture of 0.4 and a working distance of 3.9 mm (L1) to tightly focus the pulses onto a spot size below 2 µm on the magnetic sample.

Exploiting two different spectral bands for the optical probes and separate balanced detection arms has a distinct advantage compared to the ultrafast SNS scheme demonstrated previously (ref.[9]) wherein a single balanced detector receives two femtosecond optical pulses. When such a pulse impinges on a photodetector, a finite dead time in the order of tens of picoseconds emerges due to the persisting photo carriers. This nonlinearity deteriorates the detected signals when the temporal separation of the two probe pulses is small, thus critically limiting the probing window. By spectrally separating the optical pulses and detecting them individually in separate photodiodes, this problem is avoided and the ultrafast correlation dynamics can be correctly measured down to femtosecond time delays.

### b. Optical probing of the ultrafast magnetisation fluctuations

Upon transmission at times $t$ and $t+\Delta t$, transient magnetisation fluctuations $\delta M_z(t)$ parallel to the propagation direction of the beams ($z$-direction) are encoded as polarisation rotation noise of the light

$$\delta\alpha(t) \propto d\cdot\mu_0\cdot V(\lambda)\cdot\delta M_z(t)$$

(1)

via the magneto-optic Faraday effect[18]. Here, $d$ is the thickness of the sample, $\mu_0$ is the vacuum permeability, and $V(\lambda)$ is the Verdet constant as a function of wavelength $\lambda$.

After transmission, a 20 mm focal length lens (L2) collimates the beams, which then are guided into two polarimetric detector arms using a dichroic mirror (DM). Each consists of a halfwave plate (HWP), a Wollaston prism (WP) and a pair of balanced photodetectors (BPD). The WP splits the beam in its s- and p-polarised components[19], which are subsequently focused onto and detected by the BPD. The BPD yields a polarisation angle $\alpha(t)$ dependent difference current

$$\Delta I(\alpha(t)) = I_s - I_p = sP_0\left(1 - 2\cos^2(\alpha(t))\right),$$

(2)

where $s$ is the quantum efficiency of the detectors, $P_0$ the average laser power, and $I_s$ and $I_p$ are the photocurrents produced by the s- and p-polarised components of the input beam. Note that the time-dependent polarisation angle $\alpha(t) = \langle\alpha\rangle + \delta\alpha(t)$ consists of a static polarisation component $\langle\alpha\rangle$ arising, e.g., from the initial laser polarisation or the static magnetisation components along the propagation direction of light, as well as a time-dependent fluctuation part $\delta\alpha(t)$ which contains the transient magnetisation fluctuations of the sample introduced in equation (1). We use the HWP to rotate the static polarisation component of the beams to $\langle\alpha\rangle = \frac{\pi}{4} + \alpha'$, where $\alpha'$ is a small static offset caused by, e.g., imperfect balancing. In the case of $\alpha' \ll \delta\alpha(t) \ll 1$, equation (2) reduces to

$$\Delta I(\alpha(t)) \approx 2sP_0\delta\alpha(t) \propto \delta M_z(t),$$

(3)

indicating that the balanced output is directly proportional to the magnetisation fluctuation. In practice the offset angle $\alpha' \gg \delta\alpha(t)$ and the fluctuations under investigation are consequently dominated by a large static background. Furthermore, additional sources of unwanted background noise arising in the detection circuit impair the measurement, namely the photon shot noise of the probe pulses, as well as Johnson noise and $\frac{1}{f}$ noise. To extract the transient spin noise from such a large spurious background, we use two independent polarimetric detector arms, and calculate their cross correlation in a later step. By this method, uncorrelated background noise is eliminated and only the correlated noise due to the magnetisation fluctuations remains.

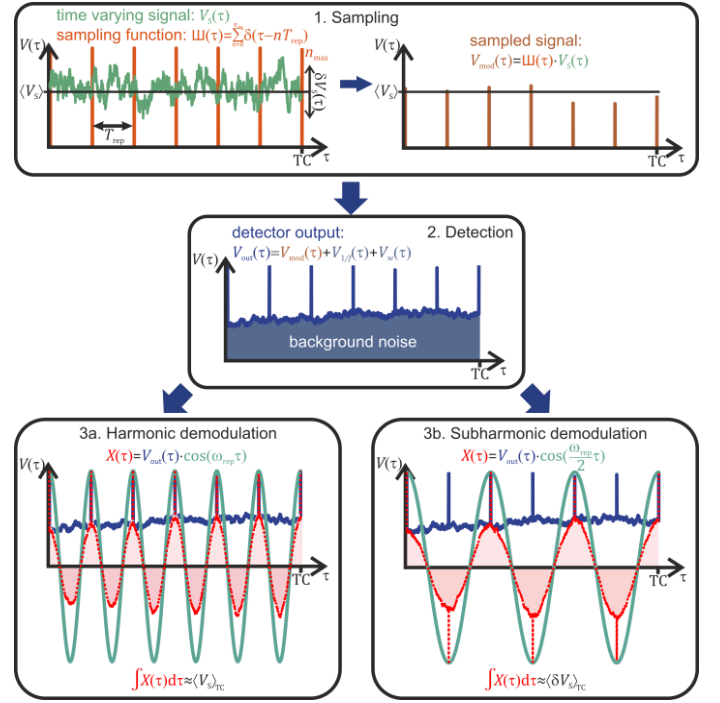

Figure 2: Difference between harmonic and subharmonic demodulation of a sampled time-varying signal $V_s(\tau) = \langle V_s\rangle + \delta V_s(\tau)$. $V_s(\tau)$ consists of a mean value $\langle V_s\rangle$ and a fluctuation part $\delta V_s$. In a first step (1. Sampling), $V_s(\tau)$ is sampled with a sampling function $\text{Ш}(\tau)$, which consists of $n_{max}$ periodic Dirac pulses equally spaced by the repetition time $T_{rep}$ within the observation interval $[0, TC]$. This yields the sampled or modulated time trace $V_{mod}(\tau)$ and emulates ultrafast probing of a fluctuating material using a pulse train of femtosecond laser pulses. Upon detection of this modified pulse train (2. Detection), $V_{mod}(\tau)$ is afflicted by $\frac{1}{f}$ – noise $V_{1/f}(\tau)$ and white background noise $V_w(\tau)$, respectively. The detector output $V_{out}(\tau) = V_{mod}(\tau) + V_{1/f}(\tau) + V_w(\tau)$ then serves as the input for the demodulation process in a lock-in amplifier. Here, $V_{out}(\tau)$ is multiplied with a reference cosine function with the pulse train frequency $\omega_{rep} = \frac{2\pi}{T_{rep}}$ to yield the mean value $\langle V_s\rangle_{TC}$ of the investigated time-varying signal $V_s(\tau)$ in the observation window $[0, TC]$ upon integration of the product (3a. Harmonic demodulation). Alternatively, multiplication with a cosine function of half the pulse train frequency $\frac{\omega_{rep}}{2}$ and subsequent integration grants access to the fluctuation part $\langle\delta V_s(\tau)\rangle_{TC}$ (3b. Subharmonic demodulation).

### c. Mathematical formulation of the subharmonic lock-in detection and subsequent correlation analysis

In the following, we first mathematically model the process of lock-in detection for general demodulation frequencies. We then focus on the case of subharmonic demodulation at half the repetition frequency $\frac{f_{\text{rep}}}{2}$, and compare it to the conventional harmonic demodulation at $f_{\text{rep}}$. In particular, we show that the subharmonic demodulation grants access to the fluctuation part $\delta V_s(\tau)$ of a time-varying signal $V_s(\tau) = \langle V_s \rangle + \delta V_s(\tau)$, where $\tau$ is general laboratory time, while removing the mean value $\langle V_s \rangle$ and $\frac{1}{f}$ noise components. In contrast, harmonic demodulation reveals $\langle V_s \rangle$ through averaging of $V_s(\tau)$ and thereby diminishing the fluctuation part. Lastly, the subsequent correlation analysis is discussed, which yields the correlated noise and removes any uncorrelated background.

**Lock-in detection of pulse trains**

[9]Given their short pulse duration, we model the femtosecond pulsed laser train by a sequence of Dirac delta functions, establishing a time-domain Dirac comb periodic with the sampling time $T_{\text{rep}} = \frac{1}{f_{\text{rep}}}$:

$$\text{III}(\tau) = \sum_{n=-\infty}^{\infty} \delta(\tau - nT_{rep}). \quad (4)$$

$\text{III}(\tau)$ denotes the sampling function and describes the arrival time of the optical probes impinging on the sample with periodicity $T_{\text{rep}}$ at laboratory time $\tau$ and $n \in \mathbb{Z}$ as the pulse number. Sampling of a time-varying signal $V_s(\tau)$ in the laboratory time frame $\tau$ with a femtosecond pulse train can therefore be approximated by simple multiplication with the sampling function (see Figure 2, panel 1):

$$V_{\text{mod}}(\tau) = V_s(\tau) \cdot \text{III}(\tau) \quad (5)$$

Upon measurement, this modulated signal $V_{\text{mod}}(\tau)$ is subject to $\frac{1}{f}$ and white noise components arising in the detection electronics (Figure 2, panel 2). We denote these quantities as $V_{1/f}(\tau)$ and $V_w(\tau)$, respectively. Consequently,

$$V_{\text{out}}(\tau) = V_{\text{mod}}(\tau) + V_{1/f}(\tau) + V_w(\tau) \quad (6)$$

is the total output of the detectors and serves as the input for the lock-in amplifier. In the first step, the lock-in performs a dual-phase demodulation[20] on this input: $V_{\text{out}}(\tau)$ is multiplied with a sinusoidal reference function $V_{d,X}(\tau) = A \cdot \cos(\omega_d \tau + \theta)$ and a 90° phase-shifted copy $V_{d,Y}(\tau) = A \cdot \sin(\omega_d \tau + \theta)$, respectively. In-phase and orthogonal quadrature components

$$X(\tau) = V_{\text{out}}(\tau) \cdot V_{d,X}(\tau) = V_{\text{out}}(\tau) \cdot [A \cdot \cos(\omega_d \tau + \theta)]$$

and

$$Y(\tau) = V_{\text{out}}(\tau) \cdot V_{d,Y}(\tau) = V_{\text{out}}(\tau) \cdot [A \cdot \sin(\omega_d \tau + \theta)] \quad (7)$$

result. Here, $A$ is the amplitude, $\omega_d$ is the angular frequency and $\theta$ is the relative phase of the reference signal. Mathematically, this mixing can also be described as multiplication with a single complex Euler's function:

$$Z(\tau) = X(\tau) + iY(\tau) = V_{\text{out}}(\tau) \cdot A e^{i(\omega_d \tau + \theta)} \quad (8)$$

Subsequently, low-pass filtering of the input is performed on the mixed signal $Z(\tau)$ in the time interval $[0, TC]$, which we approximate by simple integration[21]:

$$\langle Z(\tau) \rangle_{TC} = \frac{1}{TC} \int_0^{TC} V_{\text{out}}(\tau) \cdot A e^{i(\omega_d \tau + \theta)} \, d\tau$$

$$= \frac{A}{TC} \int_0^{TC} \left( V_{\text{mod}}(\tau) + V_{1/f}(\tau) + V_w(\tau) \right) e^{i(\omega_d \tau + \theta)} \, d\tau$$

$$= \frac{A}{TC} \int_0^{TC} \sum_{n=-\infty}^{\infty} \delta(\tau - nT_{rep}) V_s(\tau) \cdot e^{i(\omega_d \tau + \theta)} \, d\tau$$

$$+ \frac{A}{TC} \int_0^{TC} V_{1/f}(\tau) \cdot e^{i(\omega_d \tau + \theta)} \, d\tau + \frac{A}{TC} \int_0^{TC} V_w(\tau) \cdot e^{i(\omega_d \tau + \theta)} \, d\tau$$

$$(9)$$

Next, we exploit the sampling property of the Dirac delta distribution[22] $\int f(x)\delta(x - x_0)dx' = f(x_0)$ and choose the time constant $TC = n_{\max} T_{\text{rep}}$ to be a multiple of the interpulse distance $T_{\text{rep}}$ to evaluate equation (9):

$$\langle Z(\tau) \rangle_{TC} \approx \frac{A}{n_{\max}} \sum_{n=0}^{n_{\max}} \underbrace{V_s(nT_{\text{rep}})}_{V_{s_n}} \cdot e^{i(\omega_d n T_{\text{rep}} + \theta)}$$

$$+ \frac{A}{TC} \underbrace{\int_0^{TC} V_{1/f}(\tau) \cdot e^{i(\omega_d \tau + \theta)} \, d\tau}_{\approx V_{1/f}^*(\omega_d)} + \frac{A}{TC} \underbrace{\int_0^{TC} V_w(\tau) \cdot e^{i(\omega_d \tau + \theta)} \, d\tau}_{\approx V_w^*(\omega_d)}$$

$$= \frac{A}{n_{\max}} \sum_{n=0}^{n_{\max}} V_{s_n} \cdot e^{i(\omega_d n T_{\text{rep}} + \theta)} + \frac{A}{TC} \cdot V_{1/f}^*(\omega_d) + \frac{A}{TC} \cdot V_w$$

$$(10)$$

We define $V_{s_n} = \langle V_s \rangle + \delta V_{s_n}$ as the time-varying signal $V_s(t)$, probed by pulse number $n$. For large time constants ($TC \to \infty$), the second and third integrals of equation (10) can be interpreted as the Fourier transform of the background noise time traces evaluated at frequency $\omega = \omega_d$. Therefore, they correspond to the frequency-domain representations $V_{1/f}^*(\omega_d)$ and $V_w^*(\omega_d)$ of the $\frac{1}{f}$ and white noise, respectively. The white noise part is frequency-independent and therefore constant for all frequencies $V_w^*(\omega) \equiv V_w = \text{const}$.

**Harmonic demodulation**

In conventional, harmonic demodulation, the reference frequency is chosen to be equal to the angular repetition rate $\omega_{\text{rep}} = 2\pi f_{\text{rep}} = \frac{2\pi}{T_{\text{rep}}} = \omega_d$. Consequently, we can neglect the $\frac{1}{f}$ noise component for large repetition rates $\omega_{\text{rep}}$, because $V_{1/f}^*(\omega_d \to \infty) = 0$. Inserting this expression into equation (10) gives:

$$\langle Z(\tau) \rangle_{TC} \approx \frac{A}{n_{\max}} \sum_{n=0}^{n_{\max}} V_{s_n} \cdot e^{i(2\pi n + \theta)} + \frac{V_w A}{TC}$$

$$= \langle V_{s_n} \rangle_{n_{\max}} \cdot A e^{i\theta} + \frac{V_w A}{TC} \approx \langle V_s \rangle \cdot A e^{i\theta} + \frac{V_w A}{TC}.$$

$$(11)$$

Here, we exploited the fact that the fluctuation part $\delta V_{s_n}$ vanishes after averaging over many pulses $n_{\max}$ and therefore $\langle V_{s_n} \rangle_{n_{\max}} \approx \langle V_s \rangle$. Thus, harmonic demodulation suppresses the fluctuation part for long time constants $TC$, providing the mean of the time-varying signal $\langle V_s \rangle$ multiplied by a phase factor $A e^{i\theta}$ (Figure 2, panel 3a). Still, a non-zero white noise background $\frac{V_w A}{TC}$ persists in the case of finite $TC$.

**Subharmonic demodulation**

We now turn to the case of subharmonic demodulation. To access the fluctuation part $\delta V_s(\tau)$ of the pulse train, the lock-in input is mixed with reference sinusoidal functions with a subharmonic repetition frequency $\omega_d = \frac{\omega_{\text{rep}}}{m}$, where $m \in \mathbb{N}/\{0\}$. We will see that in this case, small time constants $TC$ are beneficial to measure $\delta V_s$. Consequently, the approximation of the background noise by a Fourier transform in equation (10), where large time constants were assumed, is no longer valid. Additionally, we assume that $\frac{1}{f}$ background fluctuations that predominantly appear below hundreds of kHz, are much slower than the fluctuations of interest here which are encoded in the optical probe pulse trains at tens of MHz repetition rates. Therefore, within the corresponding time scale of tens of nanoseconds, they can be approximated as constant ($V_{1/f}(\tau) \approx V_{1/f} = \text{const}$). Low-pass filtering this constant $\frac{1}{f}$ component therefore yields:

$$\frac{A}{TC}\int_0^{TC} V_{1/f}(\tau) \cdot e^{i(\omega_d\tau+\theta)} \, d\tau \approx \frac{AV_{1/f}e^{i\theta}}{TC}\int_0^{TC} e^{i\left(\frac{2\pi}{mT_{rep}}\right)\tau} \, d\tau$$
$$= -i\frac{AV_{1/f}me^{i\theta}}{2\pi n_{max}}\left(e^{2\pi i\frac{n_{max}}{m}} - 1\right).$$

(12)

Equation (12) shows that in order to avoid a static background arising from the $\frac{1}{f}$ noise component, the time constant must fulfil the boundary condition $TC = mkT_{rep}$, where $k = \frac{n_{max}}{m} \in \mathbb{N}$ is the number of subharmonic cycles within one time constant. If this applies, $e^{2\pi i\frac{n_{max}}{m}} - 1 = e^{2\pi ik} = 0$ and, consequently, the $\frac{1}{f}$ term can be neglected. In contrast, this does not apply to the white noise component $\frac{A}{TC}\int_0^{TC}V_w(\tau) \cdot e^{i(\omega_d\tau+\theta)} \, d\tau := V'_w(\tau)$, because of its frequency independence. Therefore, a finite white noise contribution also prevails in the case of the subharmonic demodulation.

In the following, we restrict the discussion to the first subharmonic reference frequency at half the repetition rate $\omega_d = \frac{\omega_{rep}}{2}$ ($m = 2$). As we show in more detail in the Supplementary Information SI2, this choice avoids preliminary averaging of the fluctuation part and maximises the output fluctuation amplitude. It is thus more advantageous compared to the case of larger $m$ values, in agreement with ref.[16].

Inserting $\omega_d = \frac{\omega_{rep}}{2}$ into equation (9) leads to:

$$\langle Z(\tau)\rangle_{TC} \approx \frac{Ae^{i\theta}}{n_{max}}\sum_{n=0}^{n_{max}} V_{s_n} \cdot e^{i\pi n} + V'_w(\tau)$$
$$= \frac{Ae^{i\theta}}{n_{max}}\sum_{n=0}^{\frac{n_{max}}{2}}\left(V_{s_{2n}} - V_{s_{2n+1}}\right) + V'_w(\tau)$$
$$= \frac{2Ae^{i\theta}}{n_{max}/2}\sum_{n=0}^{\frac{n_{max}}{2}}\left(\langle V_s\rangle + \delta V_{s_{2n}} - \langle V_s\rangle - \delta V_{s_{2n+1}}\right) + V'_w(\tau)$$
$$= 2Ae^{i\theta}\langle\Delta\delta V_{s_{2n}}\rangle_{\frac{n_{max}}{2}} + V'_w(\tau),$$

(13)

where $\langle\Delta\delta V_{s_{2n}}\rangle_{\frac{n_{max}}{2}} = \sum_{n=0}^{\frac{n_{max}}{2}}\left(\delta V_{s_{2n}} - \delta V_{s_{2n+1}}\right)$ is the mean pulse-to-pulse difference (pulse-to-pulse fluctuation) of $\frac{n_{max}}{2}$ pulse pairs sampled within the detection time window $[0, TC]$. From equation (13), we can see that by choosing the time constant to be $TC = 2T_{rep}$, the demodulation output directly reflects pulse-to-pulse fluctuations (Figure 2, panel 3b). Remember that $\delta V_s$ contains information of the investigated magnetisation fluctuation, whereas the white noise $V'_w(\tau)$ part is constituted by spurious background noise. However, usually $V'_w(\tau) \gg \delta V_s$, which is why an additional correlation analysis is performed to extract $\delta V_s$ from the dominant background.

**Correlation analysis**

Finally, we perform a cross-correlation analysis of the two sampling pulse trains which are detected and demodulated in identical, but separate optical and electronical channels. Channel 1 and 2 each yield the output $\langle Z_{1,2}(\tau)\rangle_{TC} \approx 2Ae^{i\theta_{1,2}}\langle\Delta\delta V_{1,2,s_{2n}}\rangle_{\frac{n_{max}}{2}} + V'_{1,2,w}(\tau)$ (see equation (13)) after the demodulation process. Here, $\theta_1$ and $\theta_2$ are the phases of pulse train 1 and 2 relative to the reference sinusoidal functions, respectively. In the next step, the real parts $\text{Re}(\langle Z_{1,2}(\tau)\rangle_{TC}) = \langle X_{1,2}(\tau)\rangle_{TC} = 2A\cos(\theta_{1,2}) \cdot \langle\Delta\delta V_{1,2,s_{2n}}\rangle_{\frac{n_{max}}{2}} + V'_{1,2,w}(\tau)$ are multiplied within the lock-in amplifier using its real-time calculation function. In the following, this function is called *arithmetic unit* (AU), following the notation of the RF lock-in used in the study. The resulting product is then integrated over a sufficiently long time window $TC_{AU}$ to obtain the correlation function between the two channels:

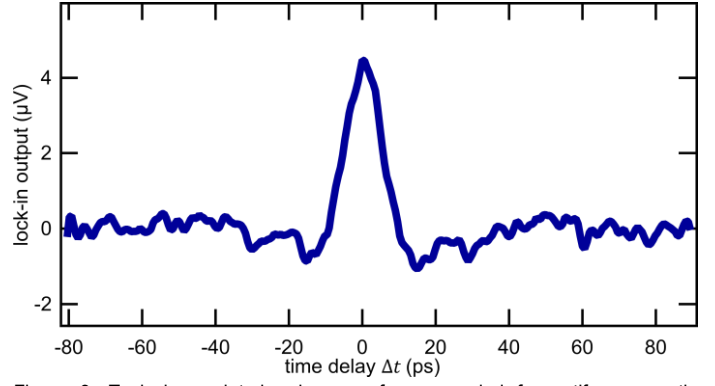

Figure 3: Typical correlated noise waveform recorded for antiferromagnetic $Sm_{0.7}Er_{0.3}FeO_3$ at 306.65 K. The correlated noise amplitude depends on the time delay $\Delta t$ of the two optical femtosecond probes.

$$\langle\langle X_1(\tau)\rangle_{TC} \cdot \langle X_2(\tau)\rangle_{TC}\rangle_{TC_{AU}}$$
$$= \frac{4A^2\cos(\theta_1)\cos(\theta_2)}{TC_{AU}}\int_0^{TC_{AU}}\left[\langle\Delta\delta V_{1,s_{2n}}\rangle_{\frac{n_{max}}{2}}\langle\Delta\delta V_{2,s_{2n}}\rangle_{\frac{n_{max}}{2}} + V'_{1,w}(\tau)V'_{2,w}(\tau)\right.$$
$$\left. + \langle\Delta\delta V_{1,s_{2n}}\rangle_{\frac{n_{max}}{2}}V'_{2,w}(\tau) + V'_{1,w}(\tau)\langle\Delta\delta V_{2,s_{2n}}\rangle_{\frac{n_{max}}{2}}\right] d\tau$$

(14)

The background noise of the distinct channels $V'_{1,2,w}(\tau)$ as well as the cross-terms between the background noise and pulse-to-pulse fluctuations $\Delta\delta V_{1,s_{2n}}$ are uncorrelated and so their correlation term vanishes for large $TC_{AU}$:

$$\int_0^{TC_{AU}} V'_{1,w}(\tau)V'_{2,w}(\tau) \, d\tau = \int_0^{TC_{AU}}\langle\Delta\delta V_{1,s_{2n}}\rangle_{\frac{n_{max}}{2}}V'_{2,w}(\tau) \, d\tau$$
$$= \int_0^{TC_{AU}} V'_{1,w}(\tau)\langle\Delta\delta V_{2,s_{2n}}\rangle_{\frac{n_{max}}{2}} \, d\tau = 0.$$

(15)

As the final output of the subharmonic demodulation and correlation analysis, we thus obtain:

$$\langle\langle X_1(\tau)\rangle_{TC} \cdot \langle X_2(\tau)\rangle_{TC}\rangle_{TC_{AU}}$$
$$= \frac{4A^2\cos(\theta_1)\cos(\theta_2)}{TC_{AU}}\int_0^{TC_{AU}}\langle\Delta\delta V_{1,s_{2n}}\rangle_{\frac{n_{max}}{2}}\langle\Delta\delta V_{2,s_{2n}}\rangle_{\frac{n_{max}}{2}} \, d\tau$$
$$= 4A^2\cos(\theta_1)\cos(\theta_2)\left\langle\langle\Delta\delta V_{1,s_{2n}}\rangle_{\frac{n_{max}}{2}}\langle\Delta\delta V_{2,s_{2n}}\rangle_{\frac{n_{max}}{2}}\right\rangle_{TC_{AU}}$$
$$= 4A^2\cos(\theta_1)\cos(\theta_2)\langle\text{CCR}_{\Delta\delta V}\rangle_{TC_{AU}}.$$

(16)

Here, $\langle\text{CCR}_{\Delta\delta V}\rangle_{TC_{AU}}$ is the mean momentary cross correlation between the pulse-to-pulse fluctuation sampled with pulse trains 1 and 2. By delaying the pulse trains with a variable ultrashort time delay $\Delta t$, $\text{CCR}_{\Delta\delta V}(\Delta t)$ directly reflects the magnon noise autocorrelation in the ultrafast time scale.

## III.  RESULTS AND DISCUSSION

The previous section shows that there are several essential parameters that critically influence FemNoC signals. Here, we systematically investigate the dependences of $\text{CCR}_{\Delta\delta V}(\Delta t)$ on each of the parameters and propose a protocol to optimise the signal-to-noise ratio from both experiment and simulation. In the experiment, we measure the real part $\langle\langle X_1(\tau)\rangle_{TC} \cdot \langle X_2(\tau)\rangle_{TC}\rangle_{TC_{AU}} = 4A^2\cos(\theta_1)\cos(\theta_2) \cdot \langle\text{CCR}_{\Delta\delta V}\rangle_{TC_{AU}}$ (see equation (16)) of the correlated magnon noise in a 10 μm thick antiferromagnetic $Sm_{0.7}Er_{0.3}FeO_3$ sample[15,23] held at 294.2 K and an external magnetic field of ~28(5) mT applied along the out-of-plane direction. We sample the noise correlation function by varying the time delays of the two pulse trains from $\Delta t = -85$ ps to $\Delta t = +85$ ps in 2 ps steps. Each data point retrieved for a fixed time delay is the result of averaging $\sim\frac{TC_{AU}}{T_{rep}}$ correlation samples. Between measurements of data points, a dwell time of $2TC_{AU}$ is implemented. Unless specified otherwise, the following measurement parameters were used: $\theta_1 = 0°$, $\theta_2 = 20°$, $TC = 50$ ns, $TC_{AU} = 1.0$ s.

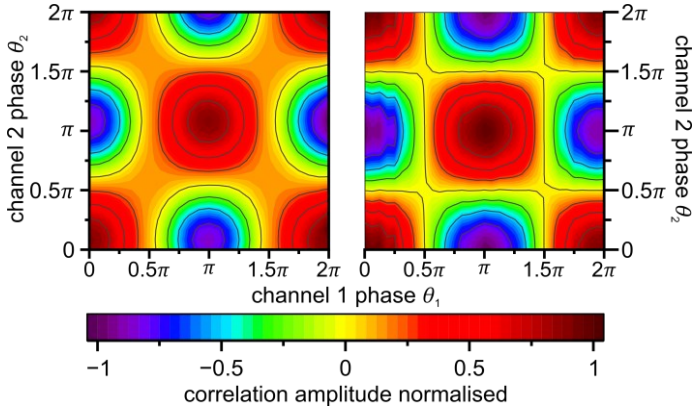

Figure 4: Normalised correlated noise amplitude as a function of the phases $\theta_1$ and $\theta_2$ of the channel 1 and 2 inputs. The experimental data (left) is recorded using the measurement parameters specified in the text, while the simulation data (right) is the result of the simulation (see Supplementary Information). The plots are normalised to the interval $[-1,1]$.

The experimental observations are compared to simulations resting on the mathematical formalism introduced in the previous Section. Details of the simulation are summarised in the Supplementary Information SI1. In the simulation, two periodic Dirac combs with a correlated amplitude modulation are prepared, thereby emulating the polarisation rotation noise imprinted on the probing pulse trains by the correlated magnon noise. Next, uncorrelated $\frac{1}{f}$ and white noise contributions are added corresponding to the spurious background arising in the detection circuit. Subsequently, the pulse trains are subject to subharmonic demodulation followed by a correlation analysis as described in the previous section. This procedure results in the correlation amplitude for a given $\Delta t$. Repeating the simulation at various time delays $\Delta t$ emulates the autocorrelation of the magnetisation obtained in the experiments.

A typical experimental waveform of the magnon correlation is depicted in Figure 3. The time trace shows a temporally symmetric function peaking at $\Delta t = 0$, followed by a gradual decrease and oscillation with a period of tens of picosecond. As we have shown in detail in ref.[15] this oscillation is ascribed to the quasi-ferromagnetic mode magnon dynamics[24], enhanced around the spin reorientation transition. It proofs that our FemNoC system can correctly retrieve the thermal fluctuations of magnetisation. In the following, we use such waveforms to verify the parameter dependences predicted by equation (16).

**Demodulation phase dependence**

First, we measure the correlation amplitude at a relative time delay of $\Delta t = 0$ (see Figure 3) for various phases $\theta_{1,2}$ of the reference sinusoidal function used in the subharmonic demodulation of the lock-in channels 1 and 2. They are varied for $2\pi$ in steps of $\frac{\pi}{12}$. Since only the real part of the correlation function (see equation ( 8 )) is measured in the experiment, the signal-to-noise is expected to strongly depend on $\theta_{1,2}$. The result is plotted in the left panel of Figure 4. It clearly shows the $\cos(\theta_1)\cos(\theta_2)$ behaviour expected from equation ( 16 ). A tiny shift of the experimental data towards positive $\theta_2$ is assigned to slightly different phases of the inputs of channels 1 and 2 arising, e.g., from minor differences in the electronic circuit. The simulation based on the introduced mathematical formalism clearly replicates the characteristics of the experimental data, as depicted in the right panel of Figure 4. As is evident from this observation, the phases of the reference to demodulate each channel need to be chosen to match the extrema positions in Figure 4 to maximise the output.

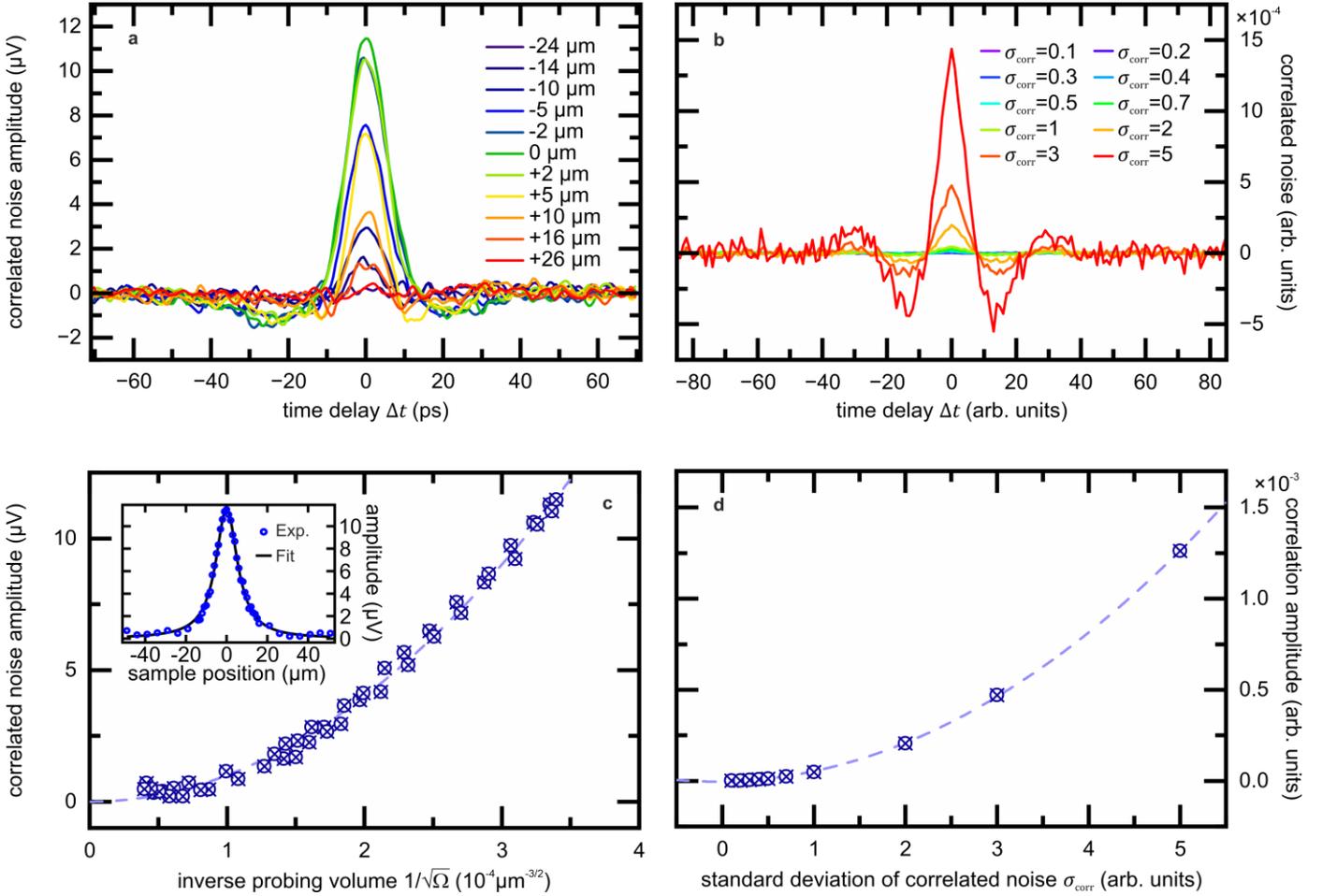

Figure 5: Correlated noise dependence on the pulse-to-pulse fluctuation amplitude. **a**. The experimental data shows the correlated noise waveforms for different longitudinal sample positions $z$ relative to the laser focus. **c**. Correlated noise amplitude for time delay $\Delta t = 0$ as a function of the square-root of the inverse probing volume $\frac{1}{\sqrt{\Omega}}$. The correlated noise clearly follows a parabolic trend as indicated by the $\frac{A}{\Omega}$ fit (purple dashed line) in accord with Ref. 15. Here, $A$ is a fitting constant. The $z$-dependent illuminated volume $\Omega(z)$ is calculated by fitting the amplitude-$z$-position dependence (inset) with the fitting function $\frac{1}{\Omega(z)} = \frac{1}{\pi w_0^2 \left[d + \frac{d^3/12 + dz^2}{z_R^2}\right]}$ (Ref. 15). Here, $d$ is the sample thickness, and $z_R$ is the Rayleigh length. **b**. The calculated waveforms are the result of the simulation (see Supplementary Information). **d**. Shows the calculated correlation amplitude for $\Delta t = 0$ as a function of the standard deviation of the correlated noise $\sigma_{corr}$. The dashed line is the result of quadratic fit of the form $A \cdot \sigma_{corr}^2$, where $A$ is a fitting constant.

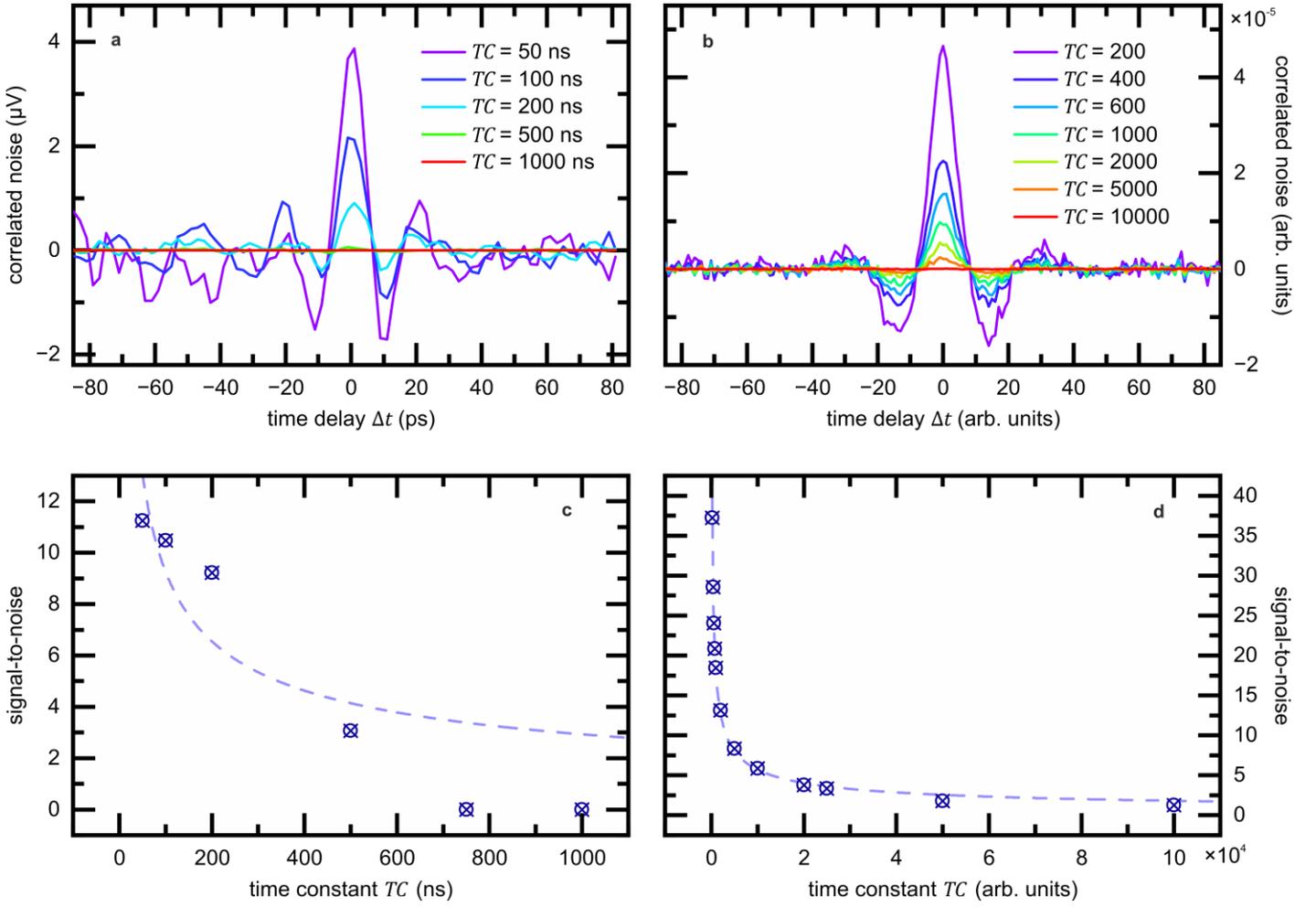

Figure 6: Correlated noise amplitude and signal-to-noise ratio as a function of the subharmonic demodulation time constant $TC$. **a**. The experimental data is recorded using the measurement parameters specified in the text, while the calculated waveforms (**b**.) are the result of the simulation (see Supplementary Information). **c**. and **d**. show the calculated signal-to-noise ratio of experiment and simulation, respectively. The dashed line is the result of an inverse square-root fit of the form $\frac{A}{\sqrt{TC}}$, where $A$ is a fitting constant.

**Amplitude dependence**

Secondly, the influence of the fluctuation amplitude on the correlated noise output is investigated. As can be seen in equation ( 16 ), the FemNoC signal is a product of the polarisation noises demodulated from two signal channels. Therefore, the output correlation signal amplitude should scale quadratically to the strength of the correlated noises. To verify this feature, we perform a numerical simulation in which the standard deviation $\sigma_{\text{corr}}$ of the simulated noisy pulse train is varied. The results are plotted in Figure 5b (see Supplementary Information for details). The correlation amplitude uniformly increases for larger amplitude of $\sigma_{\text{corr}}$. By plotting the correlation amplitude for $\Delta t = 0$ as a function of $\sigma_{\text{corr}}$ (Figure 5d), it is evident that the graph is fitted excellently by a quadratic function as expected from equation ( 16 ).

In contrast, experimentally controlling the fluctuation amplitude is not a trivial task. For this purpose, we exploit the fact that the FemNoC signal is strongly dependent on the optical probing volume on the sample. By enlarging the probe spot size, spatially incoherent magnon fluctuations are smeared out and thus the polarisation noise decreases. Specifically, polarisation noise of a single probing pulse train is expected to decrease with the square-root of the inverse probing volume $\frac{1}{\sqrt{\Omega}}$ [6]. Consequently, the correlation waveform is expected to scale as $\frac{1}{\Omega}$ [15]. We assess this aspect by studying the correlation waveform for varying longitudinal sample positions $z$ relative to the laser focus. The observed waveforms are shown in Figure 5a. The corresponding correlation amplitude at $\Delta t = 0$ ps is plotted in Figure 5c as a function of $\frac{1}{\sqrt{\Omega}}$, as well as $z$-position. The measured correlated noise amplitude drastically changes for longitudinal sample positions relative to the laser focus and ~20 μm away from the confocal position, the magnon noise is no longer visible. Its amplitude clearly follows the expected $\frac{1}{\Omega}$ dependence. This finding suggests that minimising the probing volume is crucial for enhancing the signal strength in FemNoC experiments. Tight focussing optics and careful adjustment of the sample position around the focus are warranted.

**Time constant dependence: $TC$**

We now focus on another critical parameter that influences the FemNoC signals, which is the time constant of the subharmonic demodulation $TC$. The measured correlation waveforms for various $TC$ are plotted in Figure 6a. For increasing $TC$, the absolute amplitude of the waveforms rapidly decreases. This behaviour is in good agreement with equation ( 13 ), and is interpreted as the preaveraging of the pulse-to-pulse fluctuation amplitude. At the same, it is interesting to note that the signal-to-noise ratio also decreases at larger $TC$. This observation is understood by considering the statistics of the pulse-to-pulse fluctuations: Due to their stochastic nature, averaging reduces their standard deviation as $\frac{1}{\sqrt{TC}}$. Because the fluctuation amplitude enters the correlation function once for each channel, the total amplitude dependence of the correlation output is $\text{CCR}(0) \propto \frac{1}{TC}$. In contrast, the root-mean-square (RMS) of the correlation function scales with $\sim \frac{1}{\sqrt{TC}}$, whereby the signal-to-noise ratio $\text{SNR} = \frac{\text{CCR}(0)}{\text{RMS}} \propto \frac{1}{\sqrt{TC}}$ as observed in the experimental data (Figure 6c). The simulation reproduces well the trend of the experiment, clearly following a $\frac{1}{\sqrt{TC}}$ shape (Figure 6b,d). This result suggests that indeed preaveraging of the fluctuation amplitude for large time constants is the reason for a decrease in signal-to-noise. Consequently, the time constants of the subharmonic demodulation process ideally need to be chosen to be $TC = 2T_{rep}$ to include the direct and unaveraged pulse-to-pulse fluctuation, which then serves as the input of the correlation analysis. Note that, for large time constants the simulation slightly differs from the inverse square-root fit. This finding is attributed to the calculation procedure of the SNR (see Supplementary Information SI1).

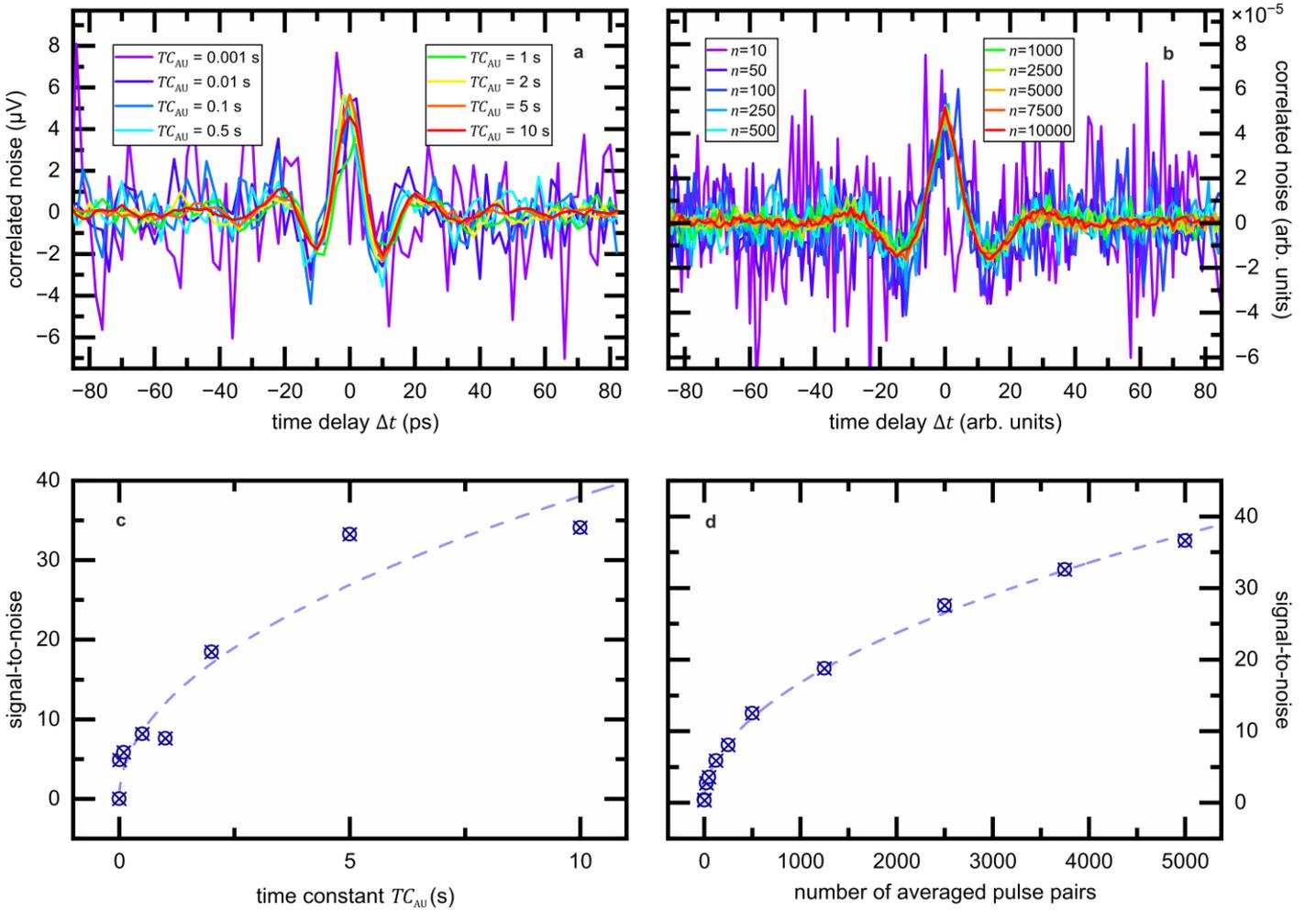

Figure 7: Correlated noise amplitude and signal-to-noise ratio as a function of the arithmetic unit's time constant $TC_{AU}$. **a.** The experimental data is recorded using the measurement parameters specified in the text, while the calculated waveforms (**b.**) are the result of the simulation (see Supplementary Information). **c.** and **d.** show the calculated signal-to-noise ratio of experiment and simulation, respectively. The dashed line is the result of square-root fit of the form $A\sqrt{TC_{AU}}$, where $A$ is a fitting constant.

**Time constant dependence: $TC_{AU}$**

At last, we investigate the influence of the time constant $TC_{AU}$ on the FemNoC signal. $TC_{AU}$ determines the time window over which the product of the subharmonic outputs is averaged. The correlated noise waveform measured for different time constants $TC_{AU}$ of the lock-in's arithmetic unit is plotted in Figure 7a. For increasing $TC_{AU}$, the waveforms become less noisy, in accord with equation (16), which can be interpreted to result from additional averaging of the pulse-to-pulse fluctuation correlation $\left\langle \langle \Delta\delta V_{1,s_{2n}}\rangle_{\frac{n_{max}}{2}} \langle \Delta\delta V_{2,s_{2n}}\rangle_{\frac{n_{max}}{2}} \right\rangle_{TC_{AU}}$ between the two channels. The signal-to-noise as a function of $TC_{AU}$ is shown in Figure 7c. The graph clearly follows a square-root trend, as expected from the averaging behaviour. The simulated autocorrelation waveforms in Figure 7b progressively become smoother by increasing the number of averaged pulse pairs, comparable to their experimental counterparts. The simulated signal-to-noise ratio (Figure 7d) also follows a distinct square-root trend. This finding shows that $TC_{AU} \to \infty$ is beneficial for the correlation-based spin noise spectroscopy measurements. Nevertheless, large $TC_{AU}$ come with the trade-off of an increased measurement time and eventual drifts need to be considered for selecting an optimum solution in any experimental implementation.

## IV. CONCLUSION

In summary, we present a thorough technical description of subharmonic demodulation-based femtosecond noise correlation spectroscopy (FemNoC). We introduce a mathematical representation of the FemNoC data acquisition scheme and determine parameters that critically influence the FemNoC measurements. The impact of the parameter values on the FemNoC signals is then verified by simulations based on a mathematical formalism and comparison to experimental data sets. At last, we identify a set of ideal measurement parameters that promise to maximise the signal-to-noise ratio of FemNoC experiments: Firstly, the relative phases $\theta_{1,2}$ of the reference function need to be carefully adjusted to enhance the output of the subharmonic demodulation. Furthermore, the absolute fluctuation amplitude is to be maximised, e.g., by carefully placing the experimental sample into a strongly focussed laser beam or choosing an appropriate pair of probe wavelengths. Lastly, to avoid preaveraging of the subharmonic component, the time constant of the demodulation $TC$ should correspond to twice the repetition rate of the laser $TC = 2T_{rep}$, whereas the time constant of the arithmetic unit should be as long as possible and limited only by practical constraints in order to increase the number of correlation samples.

## ACKNOWLEDGEMENTS

This research was supported by the Overseas Research Fellowship of the Japan Society for the Promotion of Science (JSPS), Zukunftskolleg Fellowship from the University of Konstanz, JSPS KAKENHI (JP21K14550, JP23K17748) and by the Deutsche Forschungs-gemeinschaft (DFG, German Research Foundation) — Project-ID 425217212-SFB 1432.

## AUTHOR DECLARATIONS

**Conflict of Interest**

The authors have no conflicts to disclose.

## Author Contributions

**Marvin A. Weiss:** Data curation (lead); Formal analysis (lead); Investigation (lead); Methodology (lead); Software (lead); Project Administration (lead); Writing – original draft (lead). **Franz S. Herbst:** Formal analysis (supporting); Writing – review & editing (supporting). **Stefan Eggert:** Resources (equal). **Makoto Nakajima:** Resources (equal). **Alfred Leitenstorfer:** Resources (equal); Conceptualization (equal); Writing – review & editing (supporting). **Sebastian T.B. Goennenwein:** Conceptualization (equal); Resources (equal); Writing – review & editing (supporting); Supervision (equal). **Takayuki Kurihara:** Conceptualization (equal); Investigation (supporting); Writing – original draft (supporting); Project Administration (equal); Supervision (equal).

## DATA AVAILABILITY

The data that supports the findings of this study are available from the corresponding author upon reasonable request.

# SUPPLEMENTARY INFORMATION

SI1: Simulation of correlated spin noise measurement

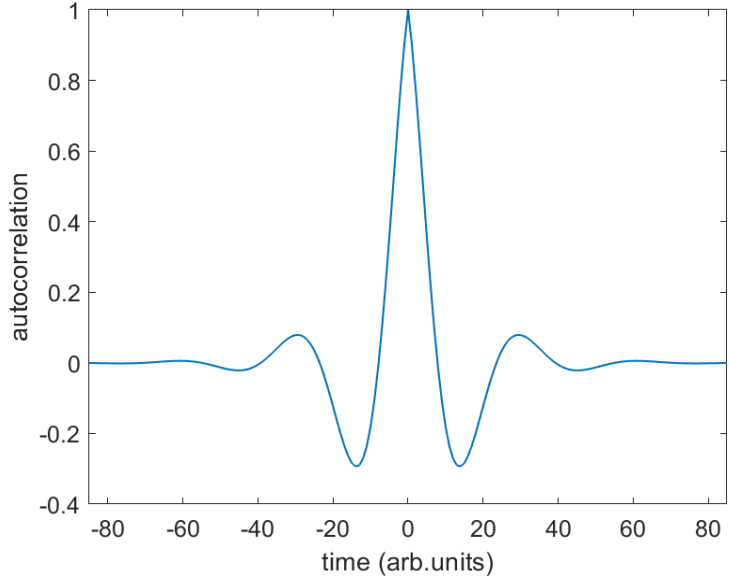

Suppl. Fig. 1: Autocorrelation as a function of arbitrary time used for the generation of correlated amplitude noise imposed on two separate Dirac combs. The autocorrelation function was generated by mirroring of an exponentially damped cosine function $e^{\left(-\frac{t}{T_{\text{corr}}}\right)} \cdot \cos\left(\frac{t}{T_{\text{mode}}}\right)$.

In the first step a reference autocorrelation function $ACR(\Delta t)$ is generated by mirroring an exponentially damped cosine function $e^{\left(-\frac{t}{T_{\text{corr}}}\right)} \cdot \cos\left(\frac{t}{T_{\text{mode}}}\right)$ in a defined range of time delays $\Delta t$ to be investigated (see Suppl. Fig. 1). Here, $T_{\text{corr}}$ denotes the time scale of the exponential damping, while $T_{\text{mode}}$ is wavelength of the cosine. Next, two Dirac combs $\text{Ш}_{1,2}(\tau) = \sum_n \delta(\tau - nT)$ (see Suppl. Fig. 2a), each consisting of $n$ Dirac pulses equally spaced by the period $T$ are created. The first Dirac comb is subject to a random amplitude modulation $z_{1_n}$ pulled from a normal distribution with variance $\sigma_{\text{corr}}$, so that $\text{Ш}'_1(\tau) = \sum_n z_{1_n} \delta(\tau - nT)$ (see Suppl. Fig. 2b). This corresponds to the undelayed pulse train impinging on the sample at times $nT$ thus probing the spin noise. Note that, we assumed $T_{\text{corr}} \ll T$, so that no correlation between consecutive probes within the same pulse train persists. The second Dirac comb $\text{Ш}_2(\tau)$ corresponds to the delayed pulse train, which probes correlated spin noise in respect to the first pulse train, depending on their relative time delay $\Delta t$. This is simulated by modulating the amplitude of $\text{Ш}_2(\tau)$ with a random variable $z'_{2_n}$. The modulated pulse train reads $\text{Ш}'_2(\tau) = \sum_n z'_{2_n} \delta(\tau - nT)$, where $z'_{2_n}$ is correlated to the amplitude modulation $z_{1_n}$ of the reference pulse train via $z'_2 = ACR(\Delta t) \cdot z_1 + \sqrt{1 - ACR^2(\Delta t)} \cdot z_2$ (ref.[25]). Additionally, random $\frac{1}{f}$ – noise $V_{1/f}(\tau)$ and white noise $V_w(\tau)$ is added to the pulse trains (see Suppl. Fig. 3). This procedure mimics the noise arising in the electrical detection of the femtosecond probes. The noise afflicted pulse trains now serve as the input of a subharmonic demodulation. Here, they are first multiplied with the subharmonic reference sinusoidal functions $\cos\left(\frac{2\pi}{m} \cdot \frac{\tau}{T_{\text{rep}}} + \theta\right)$ and $\sin\left(\frac{2\pi}{m} \cdot \frac{\tau}{T_{\text{rep}}} + \theta\right)$, where $m \in \mathbb{N}/\{0\}$ is an integer defining the order of the subharmonic number. Note that, for simplification reasons, here the arbitrary phase $\theta$ is the phase of the reference as opposed to the phase of the input signal, as introduced in the mathematical formalism of II,b. This yields the modulated pulse trains (see Suppl. Fig. 4)

$$X_{1,2}(\tau) = \left(\text{Ш}'_{1,2}(\tau) + V_{1,2_{1/f}}(\tau) + V_{1,2_w}(\tau)\right) \cdot \cos\left(\frac{2\pi}{m} \cdot \frac{\tau}{T_{\text{rep}}} + \theta\right) \text{ and } Y_{1,2}(\tau) = \left(\text{Ш}'_{1,2}(\tau) + V_{1,2_{1/f}}(\tau) + V_{1,2_w}(\tau)\right) \cdot \sin\left(\frac{2\pi}{m} \cdot \frac{\tau}{T_{\text{rep}}} + \theta\right)$$

(17)

which are subsequently integrated via the trapezoidal rule in the intervals $[kTC, (k + 1)TC]$, where $k \in \mathbb{N}_0$ to emulate low-pass filtering of the modulated signals. The results $X_{1_k}$ and $Y_{1_k}$ of demodulating pulse train 1 are then multiplied with the demodulated output of pulse train 2 $X_{2_k}$ and $Y_{2_k}$. The final result of the simulation run then yields

$$\frac{1}{k_{\max}} \sum_{k=0}^{k_{\max}} X_{1_k} X_{2_k} = \langle X_1 X_2 \rangle_k = ACR_X(\Delta t) \text{ and } \frac{1}{k_{\max}} \sum_{k=0}^{k_{\max}} Y_{1_k} Y_{2_k} = \langle Y_1 Y_2 \rangle_k = ACR_Y(\Delta t)$$

(18)

which is the autocorrelation of the $X$ and $Y$ outputs of the demodulation process for a given predefined time delay $\Delta t$. Here, $k_{\max}$ is the number of averages of the correlation. The simulation is repeated for different time delays $\Delta t$ to probe the full correlation function (see Suppl. Fig. 5). The figures shown in the main manuscript include the signal-to-noise ratio as a function of different simulation parameters. The signal-to-noise ratio of the simulation output (Suppl. Fig. 5) is calculated by first fitting the result with $A \cdot e^{\left(-\frac{t}{T_{\text{corr}}}\right)} \cdot \cos\left(\frac{t}{T_{\text{mode}}}\right)$, which is the function initially used to generate the reference autocorrelation multiplied with an amplitude term $A$. The fit is then subtracted to yield the underlying background noise still present in the correlation function, from which eventually the root-mean-square ($RMS$) is calculated. The signal-to-noise is defined as the ratio of the amplitude fit parameter and the RMS of the background noise. If $A < 1.5 \cdot RMS$, the signal-to-noise is set to zero to avoid artifacts coming from the divergence of the signal-to-noise ratio. Every signal-to-noise data point shown in the manuscript is the mean of five simulation runs.

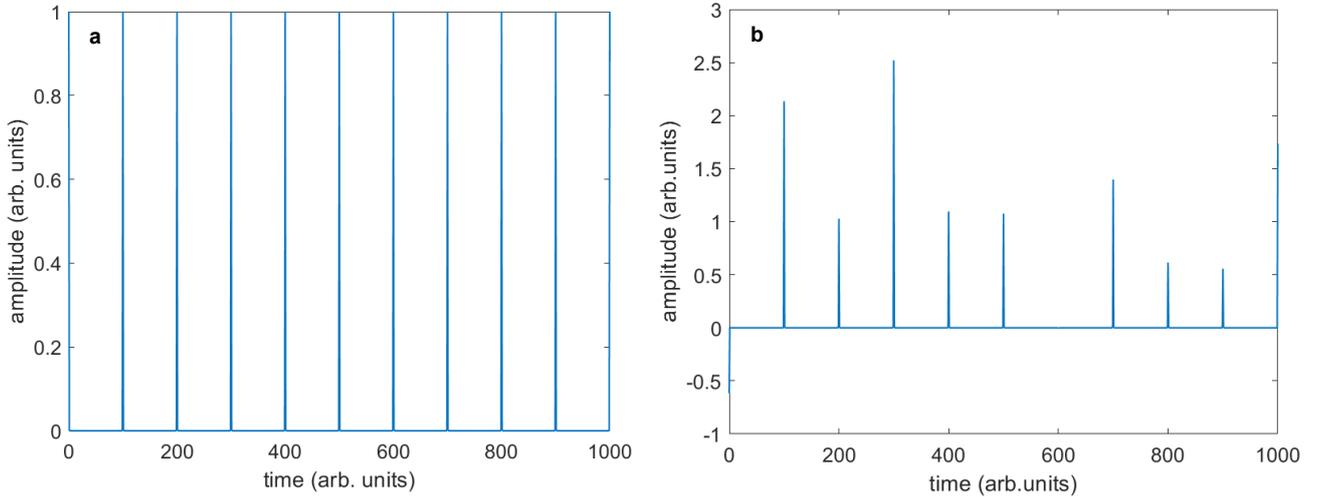

Suppl. Fig. 2: **a**, Dirac comb $\text{Ш}(\tau) = \sum_n \delta(\tau - nT)$ consisting of Dirac pulses with periodicity $T = 100$, used for the simulation of sampling spin noise with low duty-cycle femtosecond optical pulse train. **b**, Pulse train subjected to a random amplitude modulation $\text{Ш}'(\tau) = \sum_n z_n \delta(\tau - nT)$, where $z_n$ is a random variable pulled from a normal distribution with variance $\sigma_{corr}$.

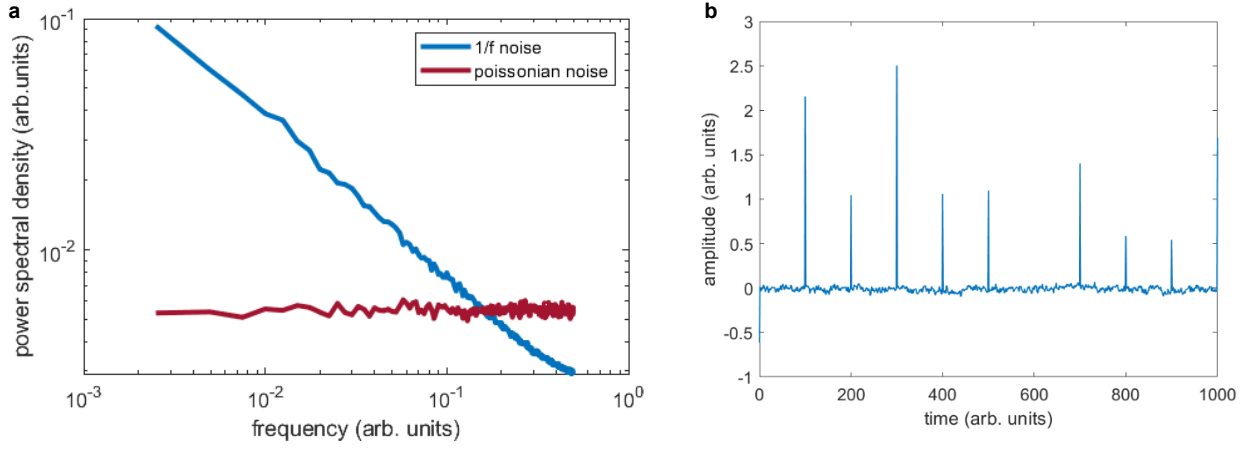

Suppl. Fig. 3: **a**, mean spectra of $\frac{1}{f}$ noise and white/Poissonian noise components, which are added to the sampling pulse trains to simulate the background noise arising in the detection circuit. **b**, amplitude-modulated and background noise afflicted pulse train $\text{Ш}'(\tau) + V_{1/f}(\tau) + V_w(\tau)$, used for the input of the subharmonic demodulation.

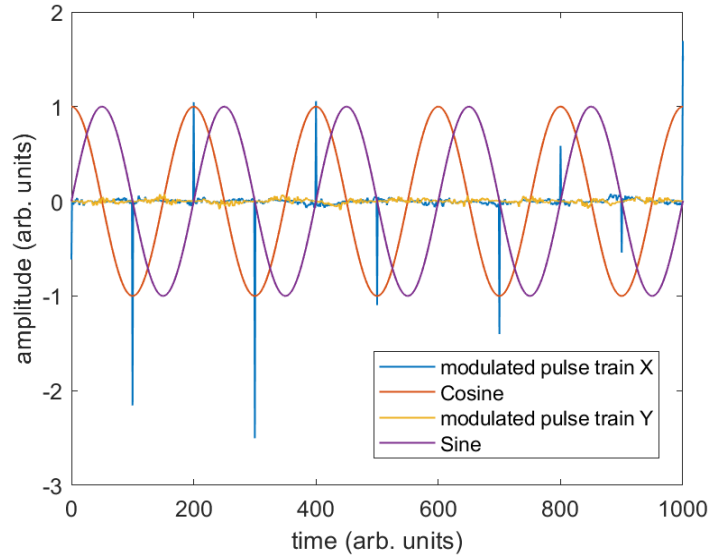

Suppl. Fig. 4: Reference sinusoidal functions $\cos\left(\frac{2\pi}{m} \cdot \frac{\tau}{T_{\text{rep}}} + \theta\right)$ (red) and $\sin\left(\frac{2\pi}{m} \cdot \frac{\tau}{T_{\text{rep}}} + \theta\right)$ (purple) used in the mixing process. Here, $m = 2$ and $\theta = 0$. The results of the mixing process are shown in blue and yellow, respectively.

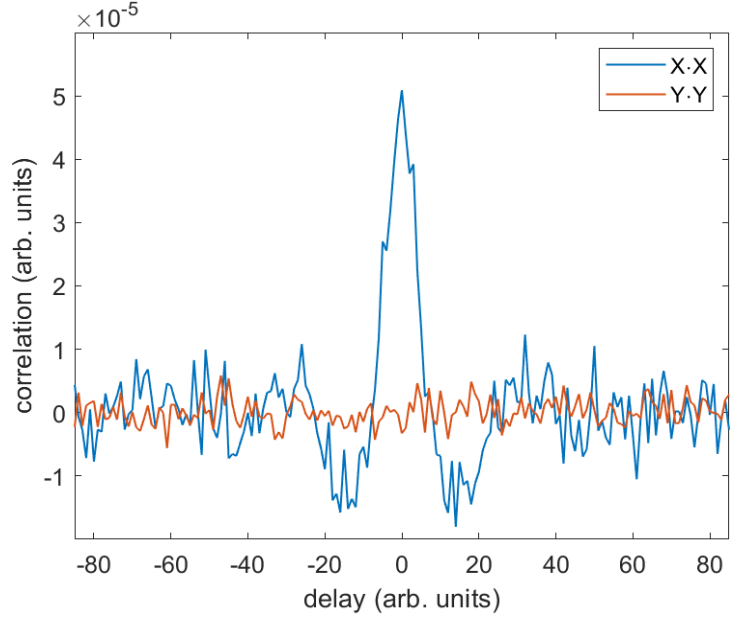

Suppl. Fig. 5: Simulated sampled autocorrelation for the subharmonic outputs $X$ (blue) and $Y$ (orange).

SI2: General derivation of mathematical formalism

Here, the demodulation process shall be described for general subharmonic reference frequencies $\omega_d = \frac{\omega_{\text{rep}}}{m}$, where, $m \in \mathbb{N}/\{0\}$ is an integer describing the order of the subharmonic frequency, whereas the mathematical formalism given in Section II,b refers to demodulation with the harmonic ($m = 1$, $\omega_d = \omega_{\text{rep}}$) and first subharmonic frequency ($m = 2$, $\omega_d = \frac{\omega_{\text{rep}}}{2}$) of the modulation frequency $\omega_{\text{rep}}$. We start our discussion from the general mixing and subsequent low-pass filtering of an input signal (equation (10)):

$$\langle Z(\tau) \rangle_{TC} \approx \frac{A}{n_{\max}} \sum_{n=0}^{n_{\max}} V_{s_n} \cdot e^{i(\omega_d n T_{\text{rep}} + \theta)} + \frac{A}{TC} \int_0^{TC} V_{1/f}(\tau) \cdot e^{i(\omega_d \tau + \theta)}\, d\tau + \frac{A}{TC} \int_0^{TC} V_w(\tau) \cdot e^{i(\omega_d \tau + \theta)}\, d\tau$$

(19)

Again, we argue that the $\frac{1}{f}$ – background fluctuations are slow compared to the other fluctuation components present in our experiment and choose $TC = mkT_{\text{rep}}$, where $k = \frac{n_{\max}}{m} \in \mathbb{N}$, so that the $\frac{1}{f}$ – term vanishes ($\frac{A}{TC} \int_0^{TC} V_{1/f}(\tau) \cdot e^{i(\omega_d \tau + \theta)}\, d\tau = 0$). Inserting $\omega_d = \frac{\omega_{\text{rep}}}{m}$ into equation (19), yields:

$$\langle Z(\tau) \rangle_{TC} \approx \frac{A e^{i\theta}}{n_{\max}} \sum_{n=0}^{n_{\max}} V_{s_n} \cdot e^{i\frac{2\pi}{m} n} + V'_w(\tau),$$

(20)

where $V'_w(\tau) = \frac{A}{TC} \int_0^{TC} V_w(\tau) \cdot e^{i(\omega_d \tau + \theta)}\, d\tau$ is a remaining white noise background. Next, we split equation (20) into a real and an imaginary part:

$$\langle Z(\tau) \rangle_{TC} \approx \frac{A e^{i\theta}}{n_{\max}} \sum_{n=0}^{n_{\max}} V_{s_n} \left[ \cos\left(\frac{2\pi}{m} n\right) + i \sin\left(\frac{2\pi}{m} n\right) \right] + V'_w(\tau)$$

(21)

To get an understanding of how equation (21) can be interpreted, we need to analyse the signs of the cosine and sine functions. Because the cosine and sine functions have the periodicity of $m$, in total $k = \frac{n_{\max}}{m}$ full periods are recorded within the total sampling time of $n_{\max} T_{\text{rep}}$. Consequently, we can split the sum from equation (21) into summation over the number of periods $k$ and $m$ pulses within a period of the subharmonic reference:

$$\langle Z(\tau) \rangle_{TC} \approx \frac{A e^{i\theta}}{n_{\max}} \sum_{j=0}^{k} \sum_{l=0}^{m} V_{s_{l_j}} \left[ \cos\left(\frac{2\pi}{m} l_j\right) + i \sin\left(\frac{2\pi}{m} l_j\right) \right] + V'_w(\tau),$$

(22)

where $l, j \in \mathbb{N}$ and $l_j$ denotes the $l$th pulse in the $j$th sampling period. The signs of the trigonometric functions change as follows:

$$\sin\left(\frac{2\pi}{m} l_j\right) \begin{cases} > 0, & \text{for } 0 < l_j < \frac{m}{2} \\ < 0, & \text{for } \frac{m}{2} < l_j < m \\ = 0, & \text{else} \end{cases} \quad \text{and} \quad \cos\left(\frac{2\pi}{m} l_j\right) \begin{cases} > 0, & \text{for } 0 < l_j < \frac{m}{4} \text{ and } \frac{3m}{4} < l_j < m \\ < 0, & \text{for } \frac{m}{4} < l_j < \frac{3m}{4} \\ = 0, & \text{else} \end{cases}$$

(23)

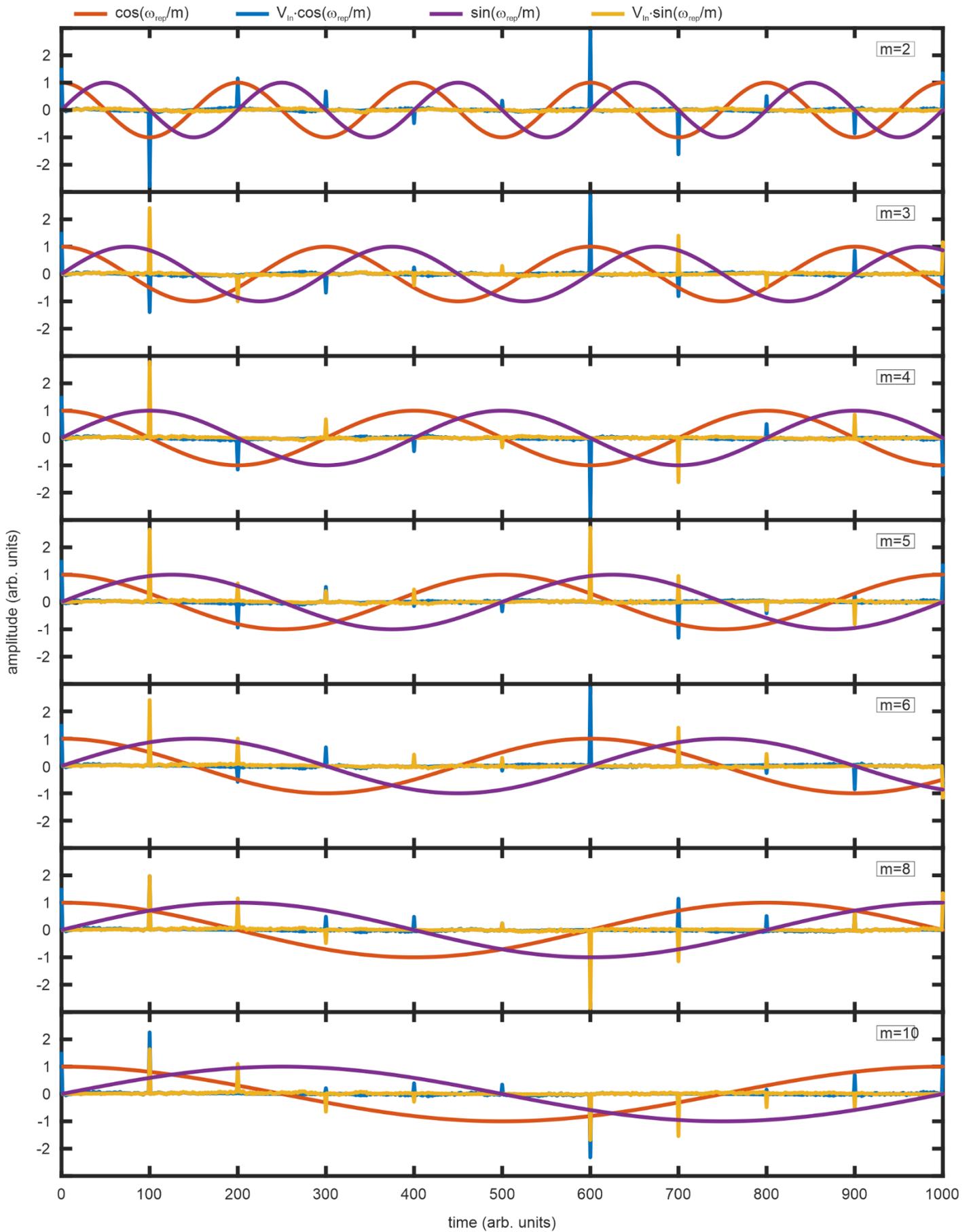

Suppl. Fig. 6: Simulated subharmonic mixing process for different subharmonic orders $m$. In the mixing process, the with frequency $\omega_{\text{rep}}$ periodic input noise-afflicted pulse train $V_{\text{in}}$ is multiplied with reference functions $\cos\left(\frac{2\pi}{m} \cdot \frac{\tau}{T_{\text{rep}}} + \theta\right)$ (red) and $\sin\left(\frac{2\pi}{m} \cdot \frac{\tau}{T_{\text{rep}}} + \theta\right)$ (purple) with frequency $\frac{\omega_{\text{rep}}}{m} = \frac{2\pi}{mT_{\text{rep}}}$ and phase $\theta = 0$. This yields the $X$ (blue) and $Y$ (yellow) outputs of the subharmonic mixing.

Using equation ( 23 ), we can further divide the sums in equation ( 22 ) into positive and negative parts:

$$\langle Z(\tau)\rangle_{TC} \approx \frac{Ae^{i\theta}}{km}\sum_{j=0}^{k}\left[\sum_{\substack{0<l_j<\frac{m}{4}\\ \frac{3m}{4}<l_j<m}}\left(\langle V_s\rangle+\delta V_{s_{l_j}}\right)\left|\cos\left(\frac{2\pi}{m}l_j\right)\right| - \sum_{\frac{m}{4}<l_j<\frac{3m}{4}}\left(\langle V_s\rangle+\delta V_{s_{l_j}}\right)\left|\cos\left(\frac{2\pi}{m}l_j\right)\right| + i\sum_{0<l_j<\frac{m}{2}}\left(\langle V_s\rangle+\delta V_{s_{l_j}}\right)\left|\sin\left(\frac{2\pi}{m}l_j\right)\right|\right.$$

$$\left. - i\sum_{\frac{m}{2}<l_j<m}\left(\langle V_s\rangle+\delta V_{s_{l_j}}\right)\left|\sin\left(\frac{2\pi}{m}l_j\right)\right|\right]$$

$$=\frac{2Ae^{i\theta}}{k}\sum_{j=0}^{k}\left[\frac{1}{m/2}\underbrace{\left(\sum_{\substack{0<l_j<\frac{m}{4}\\ \frac{3m}{4}<l_j<m}}\delta V_{s_{l_j}}\left|\cos\left(\frac{2\pi}{m}l_j\right)\right| - \sum_{\frac{m}{4}<l_j<\frac{3m}{4}}\delta V_{s_{l_j}}\left|\cos\left(\frac{2\pi}{m}l_j\right)\right|\right)}_{\mathrm{Re}\left(\Delta_{\frac{m}{2}}\delta V_s\right)}\right.$$

$$\left. + i\frac{1}{m/2}\underbrace{\left(\sum_{0<l_j<\frac{m}{2}}\delta V_{s_{l_j}}\left|\sin\left(\frac{2\pi}{m}l_j\right)\right| - \sum_{\frac{m}{2}<l_j<m}\delta V_{s_{l_j}}\left|\sin\left(\frac{2\pi}{m}l_j\right)\right|\right)}_{\mathrm{Im}\left(\Delta_{\frac{m}{2}}\delta V_s\right)}\right] = \frac{2Ae^{i\theta}}{k}\sum_{j=0}^{k}\Delta_{\frac{m}{2}}\delta V_s = 2Ae^{i\theta}\left\langle\Delta_{\frac{m}{2}}\delta V_s\right\rangle_k$$

(24)

This result can be interpreted as follows: By use of subharmonic demodulation with frequency $\omega_{\mathrm{d}} = \frac{\omega_{\mathrm{rep}}}{m}$ on average $m$ pulses are recorded within one period of a subharmonic cycle. Consequently, $\frac{m}{2}$ pulses are multiplied on average with the positive and negative intervals of the subharmonic sine and cosine function, respectively. The rearranged sum in equation (24) therefore yields the real part $\mathrm{Re}\left(\Delta_{\frac{m}{2}}\delta V_s\right)$ and imaginary part $\mathrm{Im}\left(\Delta_{\frac{m}{2}}\delta V_s\right)$ of the sine- and cosine modulated difference $\Delta_{\frac{m}{2}}\delta V_s$, which is the fluctuation between two consecutive sets containing $\frac{m}{2}$ pulses each. The result of the subharmonic demodulation is this set-to-set fluctuation averaged over the $k$ sampling periods of the reference sinusoidal functions, which is additionally modulated with a phase factor $e^{i\theta}$. if $m=2$ is inserted into equation (24), we obtain the results from the main text (equation (13)). The simulated subharmonic demodulation process for different $m$ is depicted in Suppl. Fig. 6. One can see that for increasing $m$ more pulses are recorded within each subharmonic cycle. This leads to the conclusion that by calculation of the difference between two consecutive sets containing $\frac{m}{2}$ sampling pulses pre-averaging within each of the sets occurs. This eventually leads to a reduction of the set-to-set fluctuation amplitude as can be seen in Suppl. Fig. 7. Consequently, $m=2$ results in the largest set-to-set amplitude. It should additionally be noted that for $m>2$, there exists no phase $\theta$ that exclusively contains the subharmonic output into either the real or imaginary part of the fluctuation. Instead, the set-to-set fluctuation always comprises a finite real and an imaginary part, which may further decrease the signal-to-noise ratio of the experiment depending on the experimental geometry.

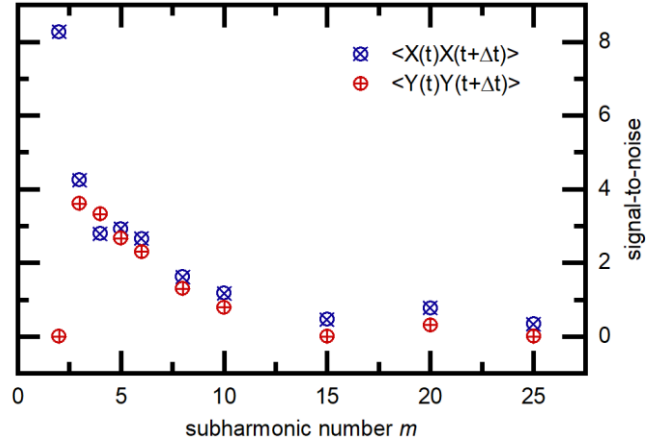

Suppl. Fig. 7: Signal-to-noise ratio of the simulated autocorrelation functions $\langle X(t)X(t+\Delta t)\rangle$ and $\langle Y(t)Y(t+\Delta t)\rangle$ for different subharmonic orders $m$ and phase $\theta = 0$.